\documentclass[prx,aps,twocolumn,superscriptaddress]{revtex4-2}

\usepackage[pdftex]{graphicx}
\usepackage{amsbsy,amssymb,amsmath,bm,mathtools}
\usepackage{amsfonts}
\usepackage{makecell}

\usepackage{xspace}
\usepackage{bm}
\usepackage{color}
\usepackage{url}
\usepackage[section]{placeins}
\usepackage[colorlinks=true,linkcolor=blue,citecolor=blue]{hyperref}

\begin{document} 

\title{Graph neural network force fields for  adiabatic dynamics of lattice Hamiltonians}
%

\author{Yunhao Fan}
\affiliation{Department of Physics, University of Virginia, Charlottesville, VA 22904, USA}

\author{Gia-Wei Chern}
\affiliation{Department of Physics, University of Virginia, Charlottesville, VA 22904, USA}

\date{\today}

\begin{abstract}
Scalable and symmetry-consistent force-field models are essential for extending quantum-accurate simulations to large spatiotemporal scales. While descriptor-based neural networks can incorporate lattice symmetries through carefully engineered features, we show that graph neural networks (GNNs) provide a conceptually simpler and more unified alternative in which discrete lattice translation and point-group symmetries are enforced directly through local message passing and weight sharing. We develop a GNN-based force-field framework for the adiabatic dynamics of lattice Hamiltonians and demonstrate it for the semiclassical Holstein model. Trained on exact-diagonalization data, the GNN achieves high force accuracy, strict linear scaling with system size, and direct transferability to large lattices. Enabled by this scalability, we perform large-scale Langevin simulations of charge-density-wave ordering following thermal quenches, revealing dynamical scaling and anomalously slow sub-Allen--Cahn coarsening. These results establish GNNs as an elegant and efficient architecture for symmetry-aware, large-scale dynamical simulations of correlated lattice systems.
\end{abstract}

\maketitle

\section{introductin}

\label{sec:intro}

Machine learning (ML) methods have revolutionized materials science by offering quantum-level accuracy at a fraction of the computational cost associated with traditional first-principles approaches~\cite{ramprasad2017,butler2018,schmidt2019,bapst2020}. This unique combination of speed, accuracy, and adaptability has significantly advanced research across diverse domains, including catalysis, energy storage, and condensed matter physics. A prominent example is the development of ML-based force-field models, which have dramatically expanded the reach of molecular dynamics (MD) simulations~\cite{behler07,bartok10,li15,behler16,shapeev16,mcgibbon17,botu17,smith17,chmiela17,chmiela18,sauceda20,zhang18,deringer19,suwa19,unke21}. By exploiting the universal function approximation capabilities of deep neural networks, these models can be trained on high-fidelity datasets from density functional theory (DFT) or beyond, accurately reproducing complex potential energy surfaces. As a result, they enable large-scale MD simulations with near {\rm ab initio} accuracy, while maintaining computational costs comparable to empirical force fields.

Two essential properties of ML force-field models are symmetry preservation and linear scalability. The former requires that surrogate ML models respect the fundamental symmetries of the underlying electronic Hamiltonians, while the latter underlies their remarkable efficiency and reflects the locality, or nearsightedness, of many-electron systems~\cite{Kohn1996,Prodan2005}. A paradigmatic framework satisfying both criteria was introduced in the pioneering works of Behler and Parrinello~\cite{behler07} and Bart\'ok {\em et al.}~\cite{bartok10}, where the total energy is partitioned into atomic contributions, $E = \sum_i \epsilon_i$, with $\epsilon_i$ denoting the atomic energy determined by the local environment of atom-$i$. Crucially, each $\epsilon_i$ is expressed as a function of a descriptor -- a set of symmetry-invariant features that encodes the local atomic environment~\cite{behler07,bartok10}. Multi-layer fully connected neural networks are commonly used to approximate this dependence, with atomic forces obtained as derivatives of the predicted total energy via automatic differentiation. 

Within the BP framework, descriptors play the key role in incorporating symmetries into the ML force-field models. For molecular systems, the relevant symmetry group is the three-dimensional Euclidean group $E(3)$, which includes translations, rotations, and reflections, together with the permutation group of atomic species. Over the years, numerous atomic descriptors have been proposed and implemented~\cite{behler07,bartok10,li15,behler11,ghiringhelli15,bartok13,himanen20,huo22}, with notable examples such as the atom-centered symmetry functions (ACSF) and group-theoretical based bispectrum coefficients. More recently, it has been shown that many of these widely used descriptors can be regarded as special cases of the Atomic Cluster Expansion (ACE) formalism~\cite{drautz19}, which offers a systematic and hierarchical framework for constructing complete, symmetry-adapted representations of atomic environments. 

The BP-type force-field framework has been extended to enable large-scale dynamical simulations of diverse condensed-matter systems, including coupled electron-spin and electron-lattice models~\cite{zhang20,zhang21,zhang22b,zhang23,cheng23a,cheng23b,Ghosh24,Fan24,tyberg25,Jang25,Liu22,Ma19}. These developments have led to the discovery of novel coarsening dynamics and phase-separation phenomena. A key distinction in such lattice-based contexts lies in the structure of the underlying symmetry groups. Specifically, the continuous Euclidean symmetry $E(3)$ of atomic systems is reduced to discrete lattice translations combined with on-site point-group symmetries. In addition, internal symmetry groups emerge that correspond to the dynamical variables themselves, such as local magnetic moments or lattice distortions. A general theoretical framework for constructing symmetry-invariant descriptors in these lattice systems has been formulated using group-theoretical methods, and several practical implementations have already been demonstrated in recent studies.

Recent advances in neural network architectures offer efficient and conceptually elegant alternatives to the BP framework. In particular, equivariant neural networks embed symmetry constraints directly into the model architecture, thereby eliminating the need for explicitly constructed symmetry descriptors~\cite{batzner2022,kaba2022,musaelian2023,gong2023,batatia2025,yang2025}. More broadly, scalable ML force-field models can be formulated within the message-passing neural network (MPNN) paradigm, which operates on the geometric and topological structure of a physical system represented as a graph~\cite{scarselli2009,gilmer2017,hamilton2017,xu2019,maron2019}. Each message-passing layer updates node or edge representations using information from a finite local neighborhood, naturally ensuring linear scalability. Moreover, symmetry preservation can be systematically enforced by requiring that symmetry-equivalent graph elements transform identically under the action of the message-passing operator. In recent years, this paradigm has spurred rapid and wide-ranging developments in ML-based force-field methodologies~\cite{xie2018,schutt2018,unke2019,choudhary2021,dai2021,reiser2022,batatia2022}, encompassing a diverse array of architectures and training strategies for {\em ab initio} molecular dynamics and materials modeling.

In this work, we introduce a message-passing neural network framework, inspired by graph neural networks (GNNs) as a special realization of MPNNs, for modeling the adiabatic dynamics of lattice Hamiltonians. The approach is designed to satisfy two essential requirements simultaneously: intrinsic scalability to large system sizes and rigorous preservation of the symmetries of the underlying Hamiltonian. The central idea is to view the lattice on which the Hamiltonian is defined as a highly symmetric graph, allowing symmetry constraints to be built directly into the network architecture rather than enforced through handcrafted features. We demonstrate this framework using the semiclassical Holstein model on a square lattice, a prototypical system for studying electron--lattice coupling. At half filling, the model exhibits a checkerboard charge-density-wave ground state that spontaneously breaks the underlying $Z_2$ sublattice symmetry.

While earlier Behler--Parrinello--type force-field models relied on carefully designed descriptors to encode the $D_4$ point-group symmetry of the square lattice, we show that the GNN framework incorporates both symmetry and scalability in a unified and conceptually transparent manner. As a result, the GNN achieves substantially improved accuracy relative to the BP approach. Moreover, large-scale thermal-quench simulations enabled by the GNN-based force field uncover unconventional domain-coarsening dynamics that lie beyond the reach of empirical phase-field descriptions, highlighting the potential of symmetry-aware neural networks for uncovering emergent nonequilibrium phenomena in lattice Hamiltonian systems.

The rest of this paper is organized as follows. In Sec.~\ref{sec:adiabatic}, we introduce the semiclassical Holstein model and outline its adiabatic lattice dynamics formulation. Section~\ref{sec:BP} briefly reviews the descriptor-based multilayer perceptron (MLP) force-field approach, providing a reference framework for scalability and symmetry preservation. Section~\ref{sec:GNN} presents the central methodological development of this work: a GNN framework for direct force prediction, in which discrete lattice translation and point-group symmetries are incorporated intrinsically through message passing. Section~\ref{sec:e-based-GNN} further extends this approach to an energy-based GNN formulation, where the network predicts local energies and the forces are obtained via automatic differentiation of the total energy, ensuring energy–force consistency. In Sec.~\ref{sec:coarsening}, we demonstrate the power of the GNN force field in enabling large-scale adiabatic Langevin dynamics simulations, focusing on the coarsening dynamics of charge-density-wave order following thermal quenches. Finally, Sec.~\ref{sec:summary} summarizes our main results and discusses future directions and broader applications of the proposed framework.

\section{Adiabatic dynamics of semiclassical Holstein model}

\label{sec:adiabatic}

Our proposed ML framework is general and applicable to dynamical simulations of a broad class of lattice systems. As a concrete proof of concept, we demonstrate its implementation and performance by studying the adiabatic dynamics of charge-density-wave (CDW) order in the Holstein model~\cite{Holstein1959}. Owing to its conceptual simplicity and its suitability for unbiased quantum Monte Carlo (QMC) simulations, the Holstein model serves as a minimal yet powerful platform for investigating electron–phonon interactions, encompassing phenomena such as polaron formation and phonon-mediated superconductivity~\cite{bonca99,golez12,mishchenko14,scalettar89,costa18,bradley21}. At half filling on bipartite lattices, including the square and honeycomb lattices, the model undergoes a finite-temperature phase transition into a commensurate CDW-ordered state~\cite{noack91,zhang19,chen19,hohenadler19}.

Despite extensive studies of CDW physics in the Holstein model, the fundamental phase-ordering dynamics of the CDW state remain computationally challenging. During relaxation toward the CDW ground state, the system is intrinsically out of equilibrium, rendering standard QMC techniques inapplicable to the direct simulation of the phase-transition dynamics. To make this problem tractable, we adopt a semiclassical approximation in which the lattice degrees of freedom are treated as classical dynamical variables. While superconducting phases require a fully quantum description, CDW order is well captured within this approximation. Indeed, the semiclassical phase diagram obtained from hybrid Monte Carlo simulations shows excellent quantitative agreement with determinant QMC results~\cite{esterlis19}.

Within this semiclassical approximation, we consider a generalized Holstein Hamiltonian $\hat{\mathcal{H}} = \hat{\mathcal{H}}_e + \mathcal{V}_L$ comprising an electronic Hamiltonian coupled to a classical elastic lattice term. The electronic part is given by
\begin{eqnarray}
	\label{eq:H_e}
	\hat{\mathcal{H}}_e = -t_{\rm nn} \sum_{\langle ij \rangle}  \hat{c}_i^\dagger  \hat{c}^{\,}_j - g \sum_i Q_i \hat{n}^{\,}_i,
\end{eqnarray}
where $\hat{c}_i^\dagger$ ($\hat{c}_i$) creates (annihilates) an spinless electron at site-$i$, $\hat{n}_i = \hat{c}^\dagger_i \hat{c}^{\,}_i$ is the electron number operator,  and $Q_i$ denotes the scalar lattice displacement at the same site.  In real materials, $Q_i$ can describe amplitudes of collective distortions such as the breathing mode of an octahedral cluster centered at site-$i$.  The first term describes nearest-neighbor hopping with amplitude $t_{\rm nn}$, while the second term represents a local electron-lattice coupling of strength $g$. The lattice distortion thus induces an on-site potential $v_i = -g Q_i$ acting on the electrons. The classical lattice energy is modeled by a system of coupled harmonic oscillators
\begin{eqnarray}
	\label{eq:V_L}
	\mathcal{V}_L = \sum_i \left( \frac{P_i^2}{2 m} + \frac{k Q_i^2}{2} \right)+ \kappa \sum_{\langle ij \rangle} Q_i Q_j. \qquad
\end{eqnarray}
where $P_i$ is the momentum conjugate to $Q_i$, $m$ is the effective ionic mass, and $k$ is the spring constant. The parameter $\kappa$ quantifies the nearest-neighbor antiferrodistortive coupling between lattice sites.

At half filling, $n_f = 1/2$, the tight-binding Hamiltonian on a square lattice exhibits perfect Fermi-surface nesting, rendering the system unstable toward a checkerboard charge modulation with ordering wave vector $\mathbf{Q} = (\pi,\pi)$ driven by the electron--lattice coupling. The resulting charge modulation is accompanied by a commensurate staggered lattice distortion of the same periodicity. This checkerboard CDW state spontaneously breaks the $Z_2$ sublattice symmetry---a special case of commensurate translational-symmetry breaking---and the associated phase transition belongs to the two-dimensional Ising universality class, consistent with QMC studies~\cite{noack91,zhang19}.

A general checkerboard CDW configuration can be characterized by a scalar order-parameter field,
\begin{eqnarray}
	\langle \hat{n}_i \rangle = \phi(\mathbf{r}_i) e^{i \mathbf{Q} \cdot \mathbf{r}_i},
\end{eqnarray}
where $\phi(\mathbf{r})$ represents the slowly varying amplitude of the local charge modulation. An analogous scalar order parameter can be introduced for the lattice distortion field~$Q_i$. Since the growth of a CDW domain can proceed through local rearrangements of electronic occupations without conserving the global order parameter, the dynamics fall into the same universality class as the nonconserved Ising order. Phenomenologically, the coarsening dynamics of the associated scalar order-parameter field are governed by model-A dynamics, also known as the time-dependent Ginzburg–Landau (TDGL) equation,
\begin{eqnarray}
	\frac{\partial \phi}{\partial t} = -\Gamma \frac{\delta \mathcal{F}}{\delta \phi} + \eta,
\end{eqnarray}
where $\Gamma$ is an empirical dissipation coefficient, $\eta(\mathbf r, t)$ is a stochastic noise field with zero mean and delta-function correlations in time, and $\mathcal{F}[\phi(\mathbf r)]$ is the effective free-energy functional,
\begin{eqnarray}
	\mathcal{F} = \int \left[ K (\nabla \phi)^2 + A \left(\phi^2 - \phi_0^2 \right)^2  \right] d^2\mathbf r.
\end{eqnarray}
Here $A$ and $K$ are positive empirical parameters, and $\phi_0$ is the equilibrium order parameter.  

The TDGL equation exhibits late-stage dynamical scaling, reflecting the self-similarity of domain morphologies at different times. In dissipative systems, the characteristic domain size follows the Allen--Cahn growth law, $L(t) \sim t^{1/2}$, arising from curvature-driven domain-wall motion. A central goal of this work is to examine whether the coarsening of CDW domains in the semiclassical Holstein model exhibits analogous scaling behavior.

To this end, we perform microscopic dynamical simulations of the lattice degrees of freedom in the adiabatic regime relevant to phase ordering. Since domain coarsening occurs on timescales much longer than electronic relaxation, the CDW evolution can be accurately captured within the adiabatic approximation, analogous to the Born--Oppenheimer scheme employed in {\em ab initio} molecular dynamics~\cite{Marx2009}. This approximation is further supported by experimental observations indicating that lattice dynamics in real CDW materials are typically one to two orders of magnitude slower than electronic dynamics~\cite{Hellmann2012,Sayers2020,Hu2022b}.

Within the adiabatic approximation, the lattice dynamics of the semiclassical Holstein model are governed by the Langevin equation,
\begin{align}\label{eqn:Langevin}
	m\frac{d^2 Q_i}{dt^2} = -\frac{\partial E}{\partial Q_i} - \gamma \frac{d Q_i}{dt} + \eta_i(t),
\end{align}
where $m$ denotes an effective ionic mass, $\gamma$ is the damping constant, and $\eta_i(t)$ is a stochastic Langevin force with zero mean and correlations
$\langle \eta_i(t) \eta_j(t') \rangle = 2 m \gamma k_B T \delta_{ij} \delta(t - t')$, ensuring relaxation toward thermal equilibrium at temperature $T$. The deterministic forces acting on the lattice degrees of freedom are obtained from the gradient of the effective energy functional,
\begin{eqnarray}
	E = \langle \hat{\mathcal{H}}_e \rangle + \mathcal{V}_L = \mathrm{Tr}(\hat{\varrho}\, \hat{\mathcal{H}}_e) + \mathcal{V}_L,
\end{eqnarray}
which consists of the electronic contribution and the classical elastic energy $\mathcal{V}_L$ associated with lattice restoring forces. Here $\hat{\varrho} = \mathcal{Z}^{-1} \exp(-\beta \hat{\mathcal{H}}_e)$ denotes the quasi-equilibrium electronic density operator within the adiabatic approximation, with $\mathcal{Z}$ the corresponding partition function. The resulting force acting on lattice site $i$, defined as $F_i = -\partial E / \partial Q_i$, can be written explicitly as
\begin{eqnarray}
	\label{eq:force1}
	F_i = g \langle \hat{n}_i \rangle - k Q_i - \kappa \textstyle \sum_{j \in \mathcal{N}(i)} Q_j,
\end{eqnarray}
where the first term arises from the electron--lattice coupling, while the remaining terms originate from the local and nearest-neighbor elastic restoring forces. The dominant computational cost in evaluating $F_i$ comes from the electronic contribution, a feature common to adiabatic dynamical simulations of lattice systems.
For the case of Holstein model, the evaluation of the forces requires computation of the local electron density which is given by 
\begin{eqnarray}
	\label{eq:ED_n}
	\langle \hat{n}_i \rangle = {\rm Tr}(\hat{\varrho} \,\hat{n}_i) = \sum_{m} f_{\rm FD}(\epsilon_m) \bigl| \phi^{(m)}_i \bigr|^2,
\end{eqnarray}
where $\epsilon_m$ and $\phi^{(m)}_i$ are the eigenenergy and eigenvector of the $m$-th single-particle eigenstate of the quadratic Hamiltonian $\hat{\mathcal{H}}_e$, respectively, and $f_{\rm FD}(\epsilon)$ denotes the Fermi--Dirac distribution. Evaluating this quantity therefore requires repeated diagonalization of the electronic Hamiltonian as the lattice evolves, motivating the development of efficient surrogate ML models for the electronic forces.

While machine-learning models—particularly neural-network–based architectures—are in principle capable of approximating arbitrary functions or mappings with high accuracy, as reviewed in Sec.~\ref{sec:intro}, their deployment as force-field models for lattice electron systems entails additional, nontrivial constraints. In the present context, a viable ML force field must satisfy two essential criteria. First, it must be scalable: models trained on finite and relatively small lattices should be directly transferable, without retraining, to simulations of much larger systems, thereby enabling access to physically relevant length scales. Second, the construction must faithfully respect the symmetries of the underlying electron–lattice Hamiltonian, ensuring that the learned forces can be derived from a symmetry-consistent energy functional and remain compatible with the original electronic problem. These requirements place strong constraints on the admissible ML architectures.

\section{Descriptor-based Multilayer perceptron approach}

\label{sec:BP}

\begin{figure*}[t]
\centering
\includegraphics[width=1.99\columnwidth]{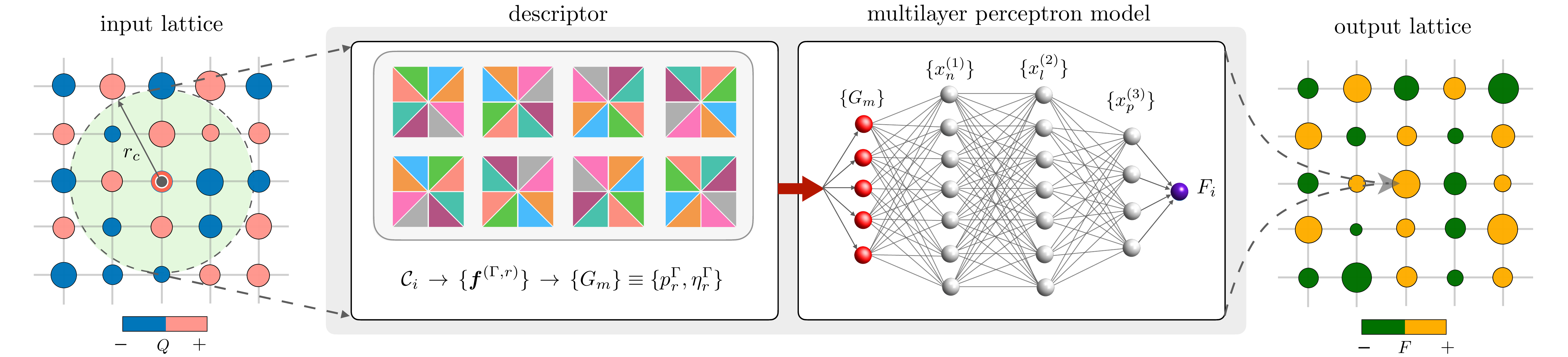}
\caption{Schematic of the BP-type force-field architecture for the semiclassical Holstein model.
Starting from the input lattice distortion configuration $\{Q_i\}$ (left), the local environment within a cutoff radius $r_c$ around a target site $i$ is first processed through a symmetry-adapted descriptor. The descriptor maps the local configuration to a set of generalized coordinates $\{G_m\}$, constructed such that all eight symmetry-related configurations generated by the $D_4$ point-group operations of the square lattice yield identical descriptor values. These symmetry-invariant descriptors $\{G_m\}$ are then fed into a fully connected neural network, which outputs a single scalar $F_i$ corresponding to the force acting on site $i$. 
}
    \label{fig:bp-scheme}
\end{figure*}

In this section, we briefly review how a scalable and symmetry-preserving ML force-field model can be constructed within the well-established Behler--Parrinello (BP) framework. This approach combines symmetry-adapted descriptors of the local environment with a multi-layer perceptron (MLP) as the learning model.  This discussion also clarifies the conceptual basis against which the GNN-based approach in the next section should be compared. As emphasized in Sec.~\ref{sec:intro}, linear scalability relies on the principle of locality. In the Holstein model, local observables such as the force $F_i$ depend only on lattice distortions within a finite spatial neighborhood of site $i$. This locality implies that the mapping from configurations to forces can be formulated entirely in terms of local environment information, allowing a model trained on relatively small systems to be transferred directly to much larger lattices without retraining.

In the original BP formulation for molecular systems, the neural network is designed to predict a local scalar energy contribution $\epsilon_i$ rather than the forces themselves. This construction is motivated by symmetry considerations: forces are vector quantities constrained by rotational and translational invariance, whereas the total energy is a scalar invariant under these transformations. The total energy is written as $E = \sum_i \epsilon_i$, where each $\epsilon_i$ depends only on the local environment of site $i$ through symmetry-adapted descriptors. Atomic forces are then obtained by differentiating $E$ with respect to the coordinates, ensuring consistency with the symmetries of the underlying Hamiltonian.

For the semiclassical Holstein model, the situation is simpler. The dynamical variables ${Q_i}$ are site-resolved scalars, and the corresponding forces are also scalars. As a result, there is no need to enforce continuous rotational covariance as in molecular systems. This permits a more direct modeling strategy in which the neural network predicts the local forces themselves, while still respecting locality and the discrete point-group symmetries of the lattice Hamiltonian. In the following, we focus on constructing such ML architectures tailored to the Holstein model.

The BP approach achieves linear scalability by explicitly exploiting locality in the construction of the ML model. As illustrated in Fig.~\ref{fig:bp-scheme}, the input to the ML model for predicting the local force $F_i$ consists of the lattice distortions within a finite neighborhood of site $i$, defined by a cutoff radius $r_c$. For convenience, we denote this local environment by
\begin{eqnarray}
	\label{eq:C_i}
	\mathcal{C}_i = \{ Q_j \, \big| \, |\mathbf r_j - \mathbf r_i| \le r_c \}.
\end{eqnarray}
The ML model is therefore trained to learn a mapping from the local configuration $\mathcal{C}_i$ to the corresponding local force $F_i$. Because the size of $\mathcal{C}_i$ is controlled solely by the fixed cutoff $r_c$ and does not grow with the total system size, the same trained model can be applied independently to every lattice site, ensuring linear scaling with system size.

To incorporate lattice point-group symmetry, the local environment is first mapped to a set of symmetry-adapted feature variables ${G_m}$. These descriptors are constructed such that two environments $\mathcal{C}_i$ and $\mathcal{C}'_i$ related by a point-group operation yield exactly the same ${G_m}$. For the square lattice with $D_4$ symmetry, the eight configurations related by discrete rotations and reflections (as shown in Fig.~\ref{fig:bp-scheme}) are therefore mapped to identical feature variables. In this way, symmetry invariance is enforced directly at the level of the input representation.

The construction of ${G_m}$ follows a systematic group-theoretical procedure. The neighborhood configuration $\mathcal{C}_i$ forms a high-dimensional reducible representation of the $D_4$ point group, which can be decomposed into irreducible representations (IRs) labeled by $\Gamma = A_1, A_2, B_1, B_2$, or $E$. As a simple example, consider the four nearest-neighbor distortions $\{ Q_a, Q_b, Q_c, Q_d \}$ surrounding a central site on a square lattice. These four degrees of freedom span a four-dimensional reducible representation, which decomposes as $4 = 1A_1 + 1B_1 + 1E$. Explicitly, the $A_1$ component corresponds to the fully symmetric combination $f^{A_1} = Q_a + Q_b + Q_c + Q_d$, the $B_1$ component corresponds to an alternating pattern such as $f^{B_1} = Q_a - Q_b + Q_c - Q_d$ (up to normalization and choice of basis), and the remaining two orthogonal combinations form the two-dimensional $E$ representation. In this symmetry-resolved basis, the local environment is expressed in terms of IR basis coefficients ${\bm f}^{(\Gamma, r)} = \bigl(f^{(\Gamma, r}_1, f^{(\Gamma, r)}_2, \cdots, f^{(\Gamma, r)}_{n_\Gamma} \bigr)$, where $n_\Gamma$ is the dimension of $\Gamma$, and $r$ enumerates the multiplicity of the IR-$\Gamma$ in the decomposition of $\mathcal{C}_i$. A first set of invariant quantities is given by the power spectrum $p^\Gamma_r = |{\bm f}^{(\Gamma, r)}|^2$, which measures the weight of the configuration in each symmetry channel.

While a complete invariant description could in principle be constructed from bispectrum coefficients that encode relative phase information between different IR sectors, this approach typically produces a large and highly redundant set of features. To obtain a compact yet symmetry-complete descriptor, we introduce reference IR coefficients ${\bm f}^{\Gamma}_{\rm ref}$ constructed from larger, symmetry-related blocks within the neighborhood $\mathcal{C}_i$. Concretely, one groups sites at a fixed radial distance or within symmetry-equivalent shells and performs the same IR decomposition on these extended blocks. Because these blocks involve many sites, the resulting ${\bm f}^{\Gamma}_{\rm ref}$ are relatively insensitive to small local fluctuations and provide a stable phase reference. The relative phase of each IR component is then defined with respect to its corresponding reference, $\eta^\Gamma_r \sim {\bm f}^{(\Gamma, r)} \cdot {\bm f}^{\Gamma}_{\rm ref}$. The final descriptor set ${G_m}$ consists of the collection of power spectra $p^\Gamma_r$ together with these phase variables $\eta^\Gamma_r$, providing a compact, nonredundant, and symmetry-invariant characterization of the local environment.

The learning model itself is implemented as a MLP, or fully connected feedforward neural network. Let ${x^{(\ell)}_m}$ denotes the feature variable associated with node $m$ in the $\ell$-th hidden layer, with the input layer defined by the symmetry-adapted descriptors characterizing the local neighborhood, i.e. $x^{(0)}_m = G_m$. The forward propagation of features through the network is given by
\begin{eqnarray}
	\label{eq:FC}
	x^{(\ell+1)}_m = \sigma\Bigl( \sum_n W^{(\ell)}_{mn} \, x^{(\ell)}_n + b^{(\ell)}_m \Bigr).
\end{eqnarray}
where $W^{(\ell)}$ and $b^{(\ell)}$ are the weight matrix and bias vector, respectively, associated with the $\ell$-th hidden layer, and $\sigma(x)$ is a nonlinear activation function such as ReLU. The nonlinear activation enables the network to capture complex, highly nontrivial dependencies of the force on the local distortions. At the output layer, the local force is obtained as a linear combination of the activations in the final hidden layer,
\begin{eqnarray}
	F_i = \sum_m W^{(D)}_m \, x^{(D)}_m + b^{(D)},
\end{eqnarray}
where $D$ denotes the total number of hidden layers. The overall ML model therefore defines a parametrized functional relationship between the local force and the symmetry-adapted descriptor of the environment:
\begin{eqnarray}
	F_i = \mathcal{F}_{\rm MLP}\left( \{G_m \}; \, \bm \theta\right), \qquad G_m = G_m(\mathcal{C}_i),
\end{eqnarray}
where $\bm\theta = \{ W, b \}$ collectively denotes the set of trainable parameters of the MLP.

\begin{table}[b]
\begin{ruledtabular}
\begin{tabular}{|c|cc|}
\textrm{Layer} & \textrm{Network} &\\
\colrule
Input (descriptor) layer &
\makecell[c]{FC$(45,4096)$\footnote{Fully Connected (FC) layer with arguments (input size, output size), corresponding to Eq.~(\ref{eq:FC}).}\\
$\sigma$\footnote{Activation function used in FC layer.} = ReLU} &\\
\hline
Hidden Layer 1 &
\makecell[c]{FC$(4096,2048)$\\ $\sigma$ = ReLU} &\\
\hline
Hidden Layer 2 &
\makecell[c]{FC$(2048,750)$\\ $\sigma$ = ReLU} &\\
\hline
Hidden Layer 3 &
\makecell[c]{FC$(750,592)$\\ $\sigma$ = ReLU} &\\
\hline
Hidden Layer 4 &
\makecell[c]{FC$(592,256)$\\ $\sigma$ = ReLU} &\\
\hline
Hidden Layer 5 &
\makecell[c]{FC$(256,127)$\\ $\sigma$ = ReLU} &\\
\hline
Hidden Layer 6 &
\makecell[c]{FC$(127,69)$\\ $\sigma$ = ReLU} &\\
\hline
Output Layer &
\makecell[c]{FC$(69,1)$ \\ $\sigma$ = Linear} &\\
\end{tabular}
\end{ruledtabular}
\caption{Architecture and hyperparameter details of the multilayer perceptron (MLP) model.}
\label{tab:bp-scheme}
\end{table}

\begin{figure}[t]
\centering
\includegraphics[width=0.99\columnwidth]{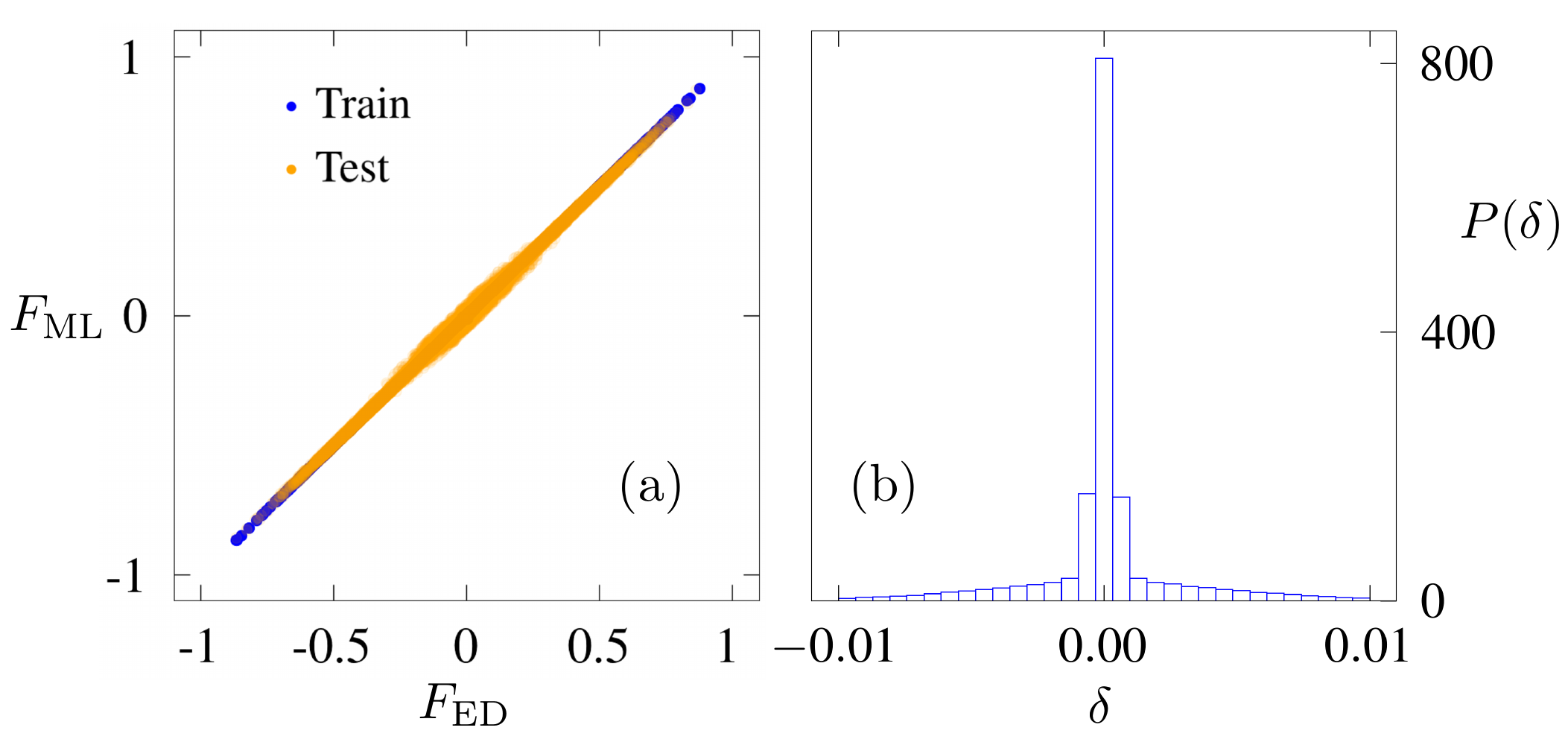}
\caption{Force prediction benchmark for the descriptor+MLP model.
(a) Parity plot comparing the machine-learning predicted force $F_{\mathrm{ML}}$ with the exact diagonalization result $F_{\mathrm{ED}}$ for both training (blue) and test (orange) datasets. The near-perfect alignment along the diagonal demonstrates excellent predictive accuracy and generalization.
(b) Distribution of the force prediction error $\delta = F_{\mathrm{ML}} - F_{\mathrm{ED}}$. The error histogram is sharply peaked around zero, indicating unbiased predictions with small variance and high numerical fidelity.}
    \label{fig:bp-benchmark}
\end{figure}

We apply this descriptor-MLP framework to the square-lattice Holstein model at half filling. The ML model is implemented in PyTorch~\cite{Paszke2019} and contains roughly $1.07\times10^{7}$ trainable parameters. The network input consists of a symmetry-aware local descriptor constructed from lattice distortions within a cutoff radius $r_c = 3.61$, chosen to capture the relevant local environment while preserving the underlying lattice symmetry. The descriptor incorporates contributions from eight neighboring shells, yielding a total of 45 input features per site. The MLP maps the resulting descriptor vector to a scalar force $F_i$ through a sequence of fully connected layers with widths 
$45 \rightarrow 4096 \rightarrow 2048 \rightarrow 750 \rightarrow 592 \rightarrow 256 \rightarrow 127 \rightarrow 69 \rightarrow 1$.
Rectified linear unit (ReLU) activations are applied after each hidden layer, while the final output layer remains linear. All weights are initialized from a normal distribution with zero mean and standard deviation 0.1, and biases are initialized from a normal distribution with zero mean and standard deviation 0.05. Further architectural details are summarized in Table~\ref{tab:bp-scheme}.

The training dataset is generated by solving the Holstein model using exact diagonalization (ED) to compute the forces according to Eqs.~(\ref{eq:force1}) and (\ref{eq:ED_n}). It comprises 200 randomly initialized lattice configurations and 400 equilibrium snapshots sampled along the full dynamical trajectory. A randomly selected $30\%$ portion of the data is held out for validation. By combining random configurations with equilibrium states, the dataset spans both generic regions of configuration space and physically relevant portions of phase space encountered during dynamics. The model is trained by minimizing a loss function defined as the mean-squared error (MSE) of the local forces,
\begin{eqnarray}
	\label{eq:loss-func}
	\mathcal{L} = \frac{1}{N} \sum_i \left| F^{\rm ML}_i - F^{\rm ED}_i \right|^2.
\end{eqnarray}
where $F^{\rm ML}_i$ and $F^{\rm ED}_i$ denote the machine-learning prediction and the ED result at site-$i$, respectively. Training is performed using the Adam optimizer~\cite{Kingma2017} with a learning rate of $10^{-3}$.

The predictive performance of the descriptor-MLP model is summarized in Fig.~\ref{fig:bp-benchmark}. Panel~(a) presents a parity plot comparing the machine-learning predicted forces $F_{\mathrm{ML}}$ with the exact diagonalization (ED) results $F_{\mathrm{ED}}$ for both the training and test datasets. The data collapse tightly onto the diagonal over the entire force range, demonstrating that the network faithfully captures the nonlinear mapping from local lattice distortions to electronic forces. The near-perfect overlap between the training (blue) and test (orange) data further indicates strong generalization and the absence of visible overfitting, despite the large number of trainable parameters. Fig.~\ref{fig:bp-benchmark}(b) shows the distribution of the force prediction error $\delta = F_{\mathrm{ML}} - F_{\mathrm{ED}}$. The error histogram is sharply peaked at zero and remains narrowly confined within a small window. Quantitatively, the standard deviation of the error is $3.06\times10^{-3}$, confirming that the predictions are both unbiased and numerically precise.

\section{Graph Neural Network Approach}

\label{sec:GNN}

\begin{figure*}[t]
\centering
\includegraphics[width=1.99\columnwidth]{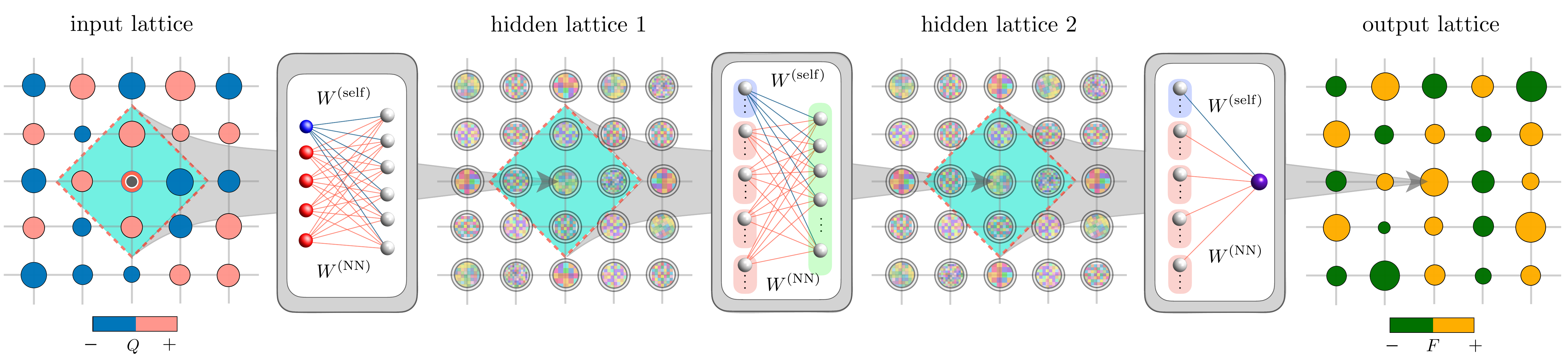}
\caption{Schematic architecture of the graph neural network (GNN) for learning the adiabatic dynamics of the Holstein model. Each layer of the network corresponds to a lattice representation of the physical system. The input layer consists of the on-site lattice distortion configuration ${Q_i}$ defined on lattice sites. At each hidden layer, the lattice sites carry multi-channel scalar node features $V^{(\ell)}_{i,\alpha}$ that encode progressively more nonlocal information through successive message-passing operations. The output layer produces the predicted forces $F_i$ acting on each lattice site. The message-passing update at each layer is illustrated schematically as a single-layer fully connected neural network acting on the local neighborhood of a site, combining same-site contributions ($W^{(\mathrm{self})}$) and neighbor contributions ($W^{(\mathrm{NN})}$) with shared parameters across the lattice. This locality-preserving and weight-sharing structure ensures translation-equivariant predictions and scalability to large lattice sizes. 
}
    \label{fig:GNN-scheme}
\end{figure*}

Graph neural networks (GNNs) are designed to learn from data that are naturally represented as networks of interconnected entities. Rather than assuming that data are organized as flattened arrays of numbers, as in a fully connected neural network, GNNs operate directly on graphs composed of nodes (which carry local features) and edges (which encode relationships or interactions). Their central mechanism is message passing: each node updates its internal representation by aggregating information from its neighbors and combining it with its own features. This aggregation is permutation-invariant, meaning that the output does not depend on how neighbors are ordered or labeled. By stacking multiple message-passing layers, a GNN progressively builds representations that incorporate information from increasingly distant parts of the graph. Because the same update rules (i.e., shared learnable parameters) are applied uniformly across all nodes and edges, GNNs naturally generalize to different system sizes and preserve relational structure.

When applied to periodic lattices, a GNN provides a particularly natural description. A lattice can be viewed as a graph in which sites are nodes and bonds define edges. Translation symmetry is automatically incorporated through weight sharing: the same message-passing operations are applied at every site, ensuring that the learned mapping treats all lattice sites equivalently. Discrete point-group symmetries are respected because symmetry-related neighborhoods correspond to relabelings of graph nodes, and the permutation-invariant aggregation guarantees that such relabelings do not change the resulting features. In this way, by treating the lattice as a graph and relying on symmetry-preserving message passing, GNNs provide an architecture that is inherently compatible with both translation symmetry and discrete lattice symmetries, without the need to explicitly construct symmetry-adapted descriptors.

From this viewpoint, GNNs on regular lattices are closely related to convolutional neural networks (CNNs), which are designed to process data defined on uniform grids. In CNNs, locality arises from the finite spatial extent of convolution kernels, while translation invariance follows from weight sharing across different sites. These properties closely parallel the local message-passing updates and shared parameters in GNNs, where node features are constructed from information in nearby neighborhoods and the same update rules are applied throughout the lattice.

An important distinction, however, concerns the treatment of lattice point-group symmetries. Standard CNN kernels typically have a fixed spatial orientation and therefore do not automatically respect discrete rotations or reflections of the lattice without additional architectural constraints or data augmentation. In contrast, within a GNN or MPNN framework, symmetry-related local environments correspond simply to permutations of neighboring nodes, and the permutation-invariant aggregation makes the incorporation of discrete lattice symmetries both natural and straightforward. This feature is particularly useful for lattice Hamiltonians such as the Holstein model, where symmetry-equivalent local configurations should be treated on equal footing.

The architecture of the GNN employed for the Holstein model is schematically illustrated in Fig.~\ref{fig:GNN-scheme}. The lattice sites are treated as nodes of a regular graph, with edges connecting nearest-neighbor sites. Since the dynamical degrees of freedom in the present problem are scalar lattice displacements, the node representations can be described entirely by scalar-valued features at the input layer, and no explicit edge features are required. Consequently, the network takes the form of a standard message-passing neural network defined on a translationally invariant lattice graph.

We denote the node features associated with site-$i$ in layer-$\ell$ by $V^{(\ell)}_{i, \alpha}$, where $\alpha$ labels the feature channels. The input layer corresponds directly to the physical lattice configuration, so that the node features are simply the scalar phonon amplitudes $V^{(1)}_i = Q_i$, with a single input channel. In the hidden layers, multiple feature channels are introduced in order to construct higher-level representations of the local lattice environment through successive nonlinear transformations.

The node representations are updated according to the standard message-passing mechanism. For each node $i$, messages are aggregated from both the node itself and its nearest-neighbor sites. The aggregated message at layer $\ell$ is defined as
\begin{eqnarray}
	\label{eq:message}
	M^{(\ell)}_{i, \alpha} = \sum_{\beta} W^{\rm (self)}_{\alpha\beta} V^{(\ell)}_{i, \beta} 
	+ \sum_{j \in \mathcal{N}(i)} \sum_\beta {W}^{\rm (NN)}_{\alpha\beta} V^{(\ell)}_{j, \beta}, \quad
\end{eqnarray}
where $\mathcal{N}(i)$ denotes the set of nearest neighbors of site $i$. The matrices $W^{\rm (self)}$ and ${W}^{\rm (NN)}$ represent trainable linear transformations applied to the self-features and neighbor features, respectively. In general, these matrices depend on the layer index $\ell$, although the superscript $\ell$ is suppressed here for notational simplicity.

This aggregation rule corresponds to a convolution-like operation on the lattice graph. The weight matrices mix information among different feature channels while remaining independent of the lattice site index. The sharing of parameters across all sites enforces lattice translation symmetry by construction. Furthermore, identical nearest-neighbor weight matrices $W^{\rm (NN)}$ are used for symmetry-equivalent neighbors, ensuring invariance under the discrete point-group symmetries of the lattice. As a result, the learned mapping depends only on the relative geometry of the lattice and not on the absolute position of a site.

Given the aggregated message $M^{(\ell)}_{i,\alpha}$, the node features at the next layer are obtained through a nonlinear update,
\begin{eqnarray}
	\label{eq:node-update}
	V^{(\ell+1)}_{i, \alpha} = \sigma\bigl( M^{(\ell)}_{i, \alpha} + b_\alpha \bigr)
\end{eqnarray}
where $\sigma(x)$ denotes a nonlinear activation function and $b_\alpha$ is a trainable bias parameter. As with the weight matrices, the bias depends only on the feature-channel index and is independent of the lattice site, thereby preserving translation invariance of the model. The nonlinear activation enables the network to represent nontrivial correlations among the lattice degrees of freedom beyond linear response, which is essential for capturing the strongly coupled electron-phonon physics of the Holstein model.

Stacking multiple message-passing layers progressively enlarges the effective receptive field of each node representation. After $\ell$ layers, the feature vector $V^{(\ell)}_{i,\alpha}$ depends on the phonon amplitudes within an $\ell$-step neighborhood of site $i$. In this way, the network constructs hierarchical representations of increasingly nonlocal lattice correlations while maintaining locality at the level of individual message-passing steps. This hierarchical structure is particularly natural for lattice Hamiltonians with short-range interactions, such as the Holstein model, where physically relevant correlations decay with distance.

Finally, the output layer maps the final node representations to the predicted on-site forces,
\begin{eqnarray}
	& & F_i = f\Bigl(V^{(D)}_{i, \alpha},  \bigr\{ V^{(D)}_{j, \alpha} \, \big| \, j \in \mathcal{N}(i) \bigr\} \Bigr) \nonumber \\
	& & \quad = \sum_\alpha W^{\rm (self)}_\alpha V^{(D)}_{i, \alpha} + \sum_{j \in \mathcal{N}(i)} W^{\rm (NN)}_\alpha V^{(D)}_{j, \alpha}, 
\end{eqnarray}
where $D$ denotes the total number of message-passing layers and $f$ represents the readout function, which in general may be linear or nonlinear. In the present implementation we adopt the simple linear readout shown in the second equality, consisting of weighted combinations of the final-layer node features on the target site and its nearest neighbors. Because the same readout function is applied uniformly across all lattice sites, the resulting mapping preserves the lattice symmetries enforced by the intermediate message-passing layers.

Successive message-passing layers progressively enlarge the spatial region over which each node representation depends on the lattice configuration. Although each update involves only nearest-neighbor communication, stacking multiple layers enables information from increasingly distant sites to be incorporated. As a result, the force at site $i$ can be expressed as a symmetry-preserving nonlinear functional of lattice distortions within a finite local environment $\mathcal{C}_i$, defined in Eq.~(\ref{eq:C_i}),
\begin{eqnarray}
	F_i = \mathcal{F}_{\rm GNN}\left( \mathcal{C}_i; \, \bm \theta \right),
\end{eqnarray}
where the spatial extent of $\mathcal{C}_i$ is determined by the depth of the network. This construction provides a hierarchical realization of a local force field consistent with the nearsightedness of lattice interactions and is therefore well suited for large-scale simulations of lattice dynamics, as illustrated in Fig.~\ref{fig:GNN-scheme}.

It is useful to contrast this formulation with the descriptor–MLP approach, where the local force depends on the environment through explicitly constructed feature variables $G_m$ that are invariant under point-group operations. In the present GNN framework, by contrast, the force $F_i$ is learned directly as a nonlinear function of the lattice distortions $Q_j$ within $\mathcal{C}_i$, with lattice symmetries enforced implicitly through the architecture rather than through predesigned descriptors.

\begin{table}[t]%
\begin{ruledtabular}
\begin{tabular}{|c|cc|}
\textrm{Layer}&\textrm{Network}&\\
\colrule
Input (embedding) layer & \makecell[c]{GCN(1,128)\footnote{A Graph Convolutional Network (GCN) layer with arguments (input size, output size) implements the message-passing and node-update operations defined in Eqs.~(\ref{eq:message}) and (\ref{eq:node-update}).}\\ $\sigma$ \footnote{The activation function used in GCN.} = ReLU} &\\
\hline
Hidden Layer 1 & \makecell[c]{GCN(128,512) \\ $\sigma$ = ReLU} &\\
\hline
Hidden Layer 2 & \makecell[c]{GCN(512,1024) \\ $\sigma$ = ReLU} &\\
\hline
Hidden Layer 3 & \makecell[c]{GCN(1024,2048) \\ $\sigma$ = ReLU} &\\
\hline
Hidden Layer 4 & \makecell[c]{GCN(2048,1024) \\ $\sigma$ = ReLU} &\\
\hline
Hidden Layer 5 & \makecell[c]{GCN(1024,512) \\ $\sigma$ = ReLU} &\\
\hline
Hidden Layer 6 & \makecell[c]{GCN(512,128) \\ $\sigma$ = ReLU} &\\
\hline
Output Layer & \makecell[c]{GCN(128,1) \\ $\sigma$ = Linear} & \\
\end{tabular}
\end{ruledtabular}
\caption{Architecture and hyperparameter details of the graph neural network (GNN) for direct force prediction of the Holstein model.}
\label{tab:GNN-scheme1}
\end{table}

The GNN model for the Holstein system is implemented in PyTorch~\cite{Paszke2019}, and a detailed specification of the architecture and hyperparameters is provided in Table~\ref{tab:GNN-scheme1}. The network consists of eight message-passing layers or graph convolutional network (GCN) layers. Each GCN layer essentially implements the updates described in Eq.~(\ref{eq:message}) and (\ref{eq:node-update}).  The first layer serves as an embedding layer that lifts the single scalar input feature into a 128-dimensional latent representation. This is followed by six hidden message-passing layers with channel dimensions $128 \rightarrow 512 \rightarrow 1024 \rightarrow 2048 \rightarrow 1024 \rightarrow 512 \rightarrow 128$, and a final output layer that produces the site-resolved force predictions. Rectified linear unit (ReLU) activation functions are used for all hidden layers, while the output layer remains linear in order to represent the continuous force values. The GNN is trained on the same dataset described in the previous section. The model parameters are optimized using the Adam optimizer~\cite{Kingma2017} with a learning rate of $10^{-3}$ by minimizing the mean-squared error between the predicted forces and the ED reference forces.

\begin{figure}[t]
\centering
\includegraphics[width=0.99\columnwidth]{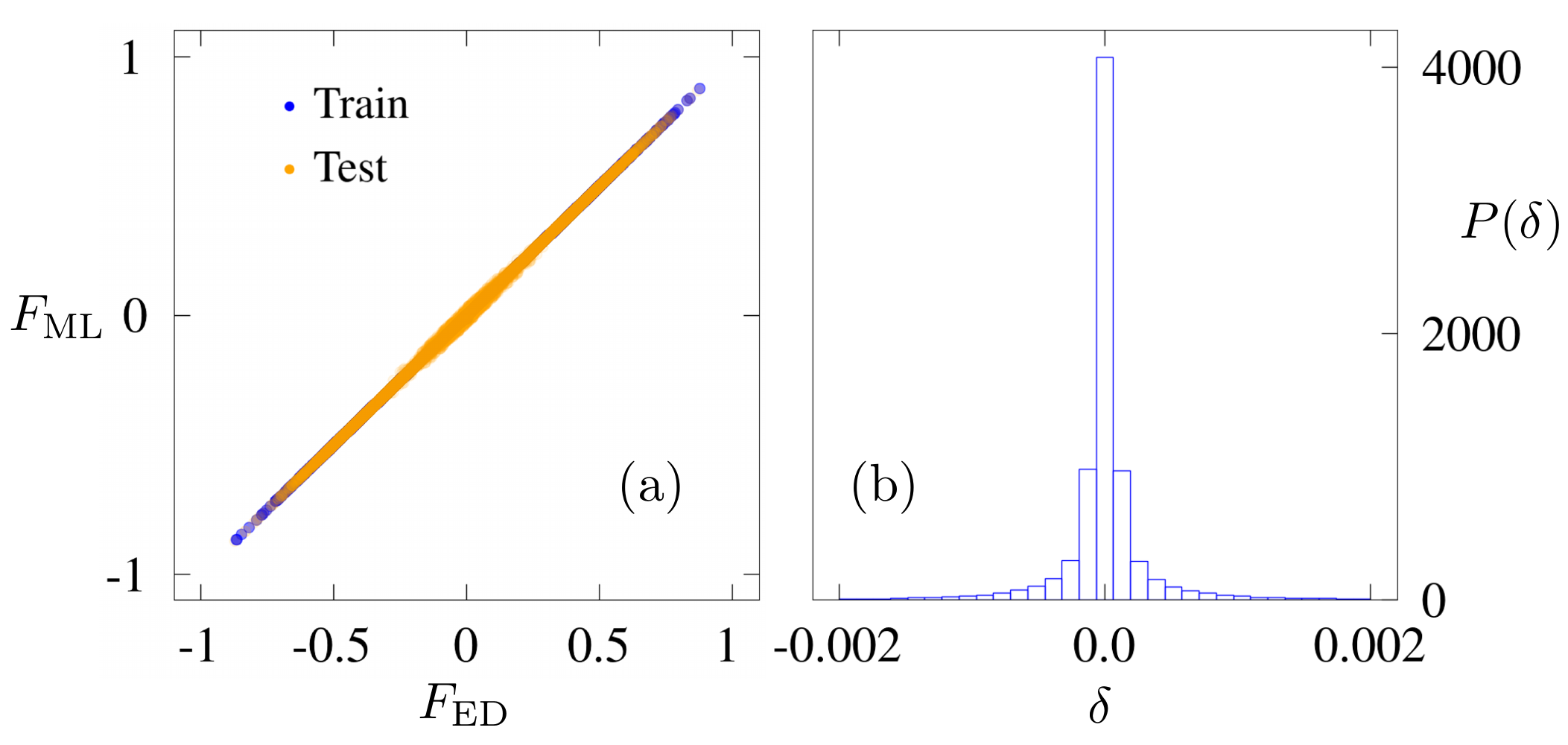}
\caption{Force prediction benchmark for the GNN model.
(a) Parity plot comparing the machine-learning predicted forces $F_{\mathrm{ML}}$ with the exact-diagonalization reference values $F_{\mathrm{ED}}$ for both training (blue) and test (orange) datasets. The close alignment of data points along the diagonal indicates high predictive accuracy and good generalization.
(b) Distribution of the force prediction error $\delta = F_{\mathrm{ML}} - F_{\mathrm{ED}}$. The error histogram is narrowly centered around zero, demonstrating unbiased predictions with small variance and high numerical accuracy.}
    \label{fig:GNN-benchmark}
\end{figure}

The predictive performance of the GNN model is summarized in Fig.~\ref{fig:GNN-benchmark}. The parity plot in Fig.~\ref{fig:GNN-benchmark}(a) compares the machine-learning predicted forces $F_{\mathrm{ML}}$ with the exact-diagonalization reference values $F_{\mathrm{ED}}$ for both the training and test datasets. The data points collapse tightly onto the diagonal over the entire range of forces, demonstrating that the network accurately reproduces the adiabatic forces and generalizes well beyond the configurations used for training. The nearly indistinguishable distributions of training and test points indicate that overfitting is minimal and that the learned force functional $\mathcal{F}_{\bm\theta}$ captures the essential dependence of the forces on the local lattice environment. 

Further insight is provided by the distribution of the force prediction error $\delta = F_{\mathrm{ML}} - F_{\mathrm{ED}}$ shown in Fig.~\ref{fig:GNN-benchmark}(b). The error histogram is sharply centered around zero with a narrow spread, indicating that the predictions are essentially unbiased and numerically precise across the dataset. The small magnitude of the residual errors confirms that the GNN provides an accurate representation of the local adiabatic force functional. These results demonstrate that the message-passing architecture is capable of learning the nonlinear dependence of the Holstein forces on the surrounding lattice distortions while preserving locality and lattice symmetries, making it well suited for large-scale dynamical simulations.

\section{Energy-based GNN model}

\label{sec:e-based-GNN} 
 
The GNN model presented in the previous section is formulated to directly predict the local forces. While this direct force-prediction strategy is effective for the Holstein model, an alternative and often more systematic approach is to construct an energy-based GNN force field, in the spirit of the original BP framework. In such formulations, the network is trained to predict site-resolved local energies, $\epsilon_i$, and the forces are subsequently obtained via automatic differentiation of the total energy with respect to the dynamical variables. This energy-based construction offers several conceptual and practical advantages.

First and foremost, it enforces energy conservation by construction: because the forces are derived from a scalar potential, they are guaranteed to be conservative. This property is essential for stable and physically consistent molecular dynamics simulations, particularly for long-time evolution where small non-conservative errors may accumulate and lead to unphysical energy drift. In contrast, direct force prediction does not automatically guarantee integrability, and additional consistency checks or constraints may be required to ensure physical reliability.

A second important advantage concerns symmetry. The direct force-prediction strategy is effective here largely because the Holstein model is simple: the dynamical variables $Q_i$ are scalar lattice distortions, and the forces are likewise scalar, making symmetry constraints straightforward to impose. In more general settings—such as systems with Jahn–Teller distortions, multi-component lattice modes, or local magnetic moments—the dynamical variables possess internal structure (e.g., vector or tensor character). The corresponding forces must then transform properly under spatial rotations, point-group operations, or other internal symmetries, making the construction of a symmetry-consistent, directly predicted force field considerably more nontrivial.

The energy-based formulation offers a modular and systematic way to incorporate symmetry. In our framework, the GNN enforces locality and discrete lattice symmetries—such as translation and point-group invariance—through its message-passing structure. Continuous or internal symmetries of the dynamical variables are incorporated at the level of the local energy functional. By using symmetry-adapted descriptors or equivariant neural-network components to predict the scalar local energy, the correct transformation properties of derived quantities follow automatically. For example, if the energy is invariant under internal rotations, its gradient with respect to a vector degree of freedom transforms covariantly, ensuring that the resulting force behaves properly. In this way, the GNN accounts for lattice symmetry and locality, while descriptor-based or equivariant constructions enforce internal symmetry, together producing a conservative and symmetry-consistent force field.

Here we present a GNN framework for predicting the local on-site energy, $\epsilon_i$. Unlike forces, which can be computed directly from exact diagonalization (ED) and used as supervised targets, the local energies are not uniquely accessible from ED. In this sense, $\epsilon_i$ should be regarded as a latent quantity within the machine-learning model, constrained only indirectly through force observables (via energy gradients). This indirect supervision renders the learning problem intrinsically more challenging and less well-conditioned than direct force prediction.

Empirically, GNNs with a capacity comparable to those used for direct force prediction perform significantly worse in the energy-based formulation. A natural remedy is to increase network depth by adding more message-passing layers. However, this is neither physically justified nor computationally efficient. Greater depth enlarges the receptive field and introduces unnecessary long-range correlations, effectively extending the model’s notion of locality beyond what the physics requires. Because distant lattice distortions $Q$ contribute only weakly to the local energy and forces, a deeper GNN yields limited accuracy gains while increasing computational cost and the risk of over-parameterization.

Instead of increasing depth, we enhance the expressive power of each propagation step. Specifically, we insert MLPs before and after message aggregation. Pre-aggregation MLPs provide richer nonlinear embeddings of local node (and edge) features, while post-aggregation MLPs enable more flexible mixing of the combined messages. This approach increases representational capacity within each layer without expanding the spatial receptive field, allowing the model to capture more intricate local energy landscapes while preserving locality and efficiency.

Specifically, we refine the node update rule beyond the simpler scheme introduced previously by introducing nonlinear feature processing prior to message aggregation. The node features at the central site $i$ and its nearest neighbors $j \in \mathcal{N}(i)$ are first passed through separate MLPs, which map them to a (potentially different) feature space and enhance their expressive representation. The transformed features are then aggregated according to
\begin{eqnarray}
	\label{eq:message2}
	& & M^{(\ell)}_{i, \alpha} = \sum_{\beta} W^{\rm (self)}_{\alpha\beta}\, \mathrm{MLP}\left( \{ V^{(\ell)}_{i, \beta} \}; \, \bm\theta^{\rm (self)} \right)  \\
	& & \qquad + \sum_{j \in \mathcal{N}(i)} \sum_\beta {W}^{\rm (NN)}_{\alpha\beta} \,\mathrm{MLP}\left( \{ V^{(\ell)}_{j, \beta} \}; \, \bm\theta^{\rm (NN)} \right), \nonumber
\end{eqnarray}
Here The matrices $W^{\rm (self)}$ and $W^{\rm (NN)}$ then linearly mix feature channels and project them into the output space indexed by $\alpha$,  $\mathrm{MLP}(\cdot, \, \bm\theta)$ denotes a multi-layer perceptron with trainable parameters $\bm{\theta}$, and distinct parameter sets are used for the central (“self”) and neighboring (“NN”) sites. The self term plays a role analogous to a self-loop in conventional message-passing networks, preserving explicit on-site information, while the NN term governs how neighboring features are weighted and combined, thereby encoding short-range correlations. By separating these contributions, the model gains additional flexibility in balancing local and inter-site effects, while translational symmetry is maintained through weight sharing across all lattice sites. Although Eq.~(\ref{eq:message2}) allows for general multi-layer MLPs with nonlinear activations, in our implementation we adopt a single fully connected layer as defined in Eq.~(\ref{eq:FC}).

The aggregated message is subsequently combined with the original node feature through a second update MLP to produce an intermediate representation:
\begin{eqnarray}
	\label{eq:update_MLP}
	U^{(\ell)}_{i, \alpha} = \sigma\!\left( M^{(\ell)}_{i,\alpha} + \sum_{\beta} W^{\rm (self, 2)}_{\alpha\beta} V^{(\ell)}_{i, \beta} + b^{\rm (self, 2)}_\alpha\right). \quad
\end{eqnarray}
Here $W^{\rm (self,2)}$ and $b^{\rm (self,2)}$ are trainable parameters of this update MLP, and $\sigma(x)$ is a nonlinear activation. This stage performs a nonlinear mixing between the original on-site state and the aggregated neighborhood message, thereby enhancing the capacity of each message-passing block without enlarging its spatial receptive field. In effect, the model increases expressivity within a fixed locality range rather than relying on additional layers to propagate information farther.

\begin{table}[t]
\begin{ruledtabular}
\begin{tabular}{|c|c|}
\textrm{Layer} & \textrm{Network} \\
\colrule
Input layer &
\makecell[c]{Node feature dimension = 1} \\
\hline
Message-passing block 1 &
\makecell[c]{
Aggregate MLP$(1,512)$\footnote{Aggregate MLP with arguments (input channels, output channels) implements Eq.~(\ref{eq:message2}).}\\
Update MLP$(1+512 \rightarrow 512)$\footnote{Update MLP with arguments (input channels, output channels) implements Eq.~(\ref{eq:update_MLP}).}\\ 
$\sigma$\footnote{The activation function used in $\mathrm{MLP}(\cdot)$ and Eq.~(\ref{eq:update_MLP}).} = ReLU\\
Residual projection$(1 \rightarrow 512)$\footnote{The residual projection implements Eq.~(\ref{eq:residual_proj}).}
} \\
\hline
Message-passing block 2 &
\makecell[c]{
Aggregate MLP$(512,1024)$\\
Update MLP$(512+1024 \rightarrow 1024)$\\
$\sigma$ = ReLU\\
Residual projection$(512 \rightarrow 1024)$
} \\
\hline
Message-passing block 3 &
\makecell[c]{
Aggregate MLP$(1024,512)$\\
Update MLP$(1024+512 \rightarrow 512)$\\
$\sigma$ = ReLU\\
Residual projection$(1024 \rightarrow 512)$
} \\
\hline
Output block &
\makecell[c]{
Readout MLP$(512+1 \rightarrow 1)$\\
$\sigma$ = ReLU\\
} \\
\end{tabular}
\end{ruledtabular}
\caption{Message-passing neural network (MPNN) architecture and hyperparameters used in this work.}
\label{tab:mpnn_parameters}
\end{table}
\footnotetext{
Each message-passing block comprises three fully connected components:
(i) a message-passing neural network (MPNN),
(ii) an update multilayer perceptron (Update MLP), and
(iii) a residual projection.
The MPNN employs a two-layer fully connected message function, while the update and residual paths are implemented as fully connected linear mappings.
}

To further facilitate stable optimization and efficient information flow across layers, we introduce an explicit residual bridge connection. The input feature is projected through a trainable linear transformation and added to the layer-normalized output of the update MLP, yielding the updated node feature:
\begin{eqnarray}
	\label{eq:residual_proj}
	V^{(\ell+1)}_{i, \alpha} = \sum_\beta W^{\rm (res)}_{\alpha\beta} V^{(\ell)}_{i, \beta} + \mathrm{LN}\bigl( U^{(\ell)}_{i, \alpha} \bigr). 
\end{eqnarray}
Here $W^{\rm (res)}$ denotes the weight matrix of the residual projection at layer $\ell$, and $\mathrm{LN}(\cdot)$ represents layer normalization. The residual pathway improves gradient flow and mitigates vanishing-gradient issues in deeper architectures, while layer normalization stabilizes training by controlling feature statistics across channels.

At the output layer, a message aggregation of the same form as Eq.~(\ref{eq:message2}) is performed. The local energy is then constructed as a linear combination of the residual (on-site) contribution and the aggregated self and neighbor messages,
\begin{eqnarray}
	& & \epsilon_i = \sum_{\alpha} W^{\rm (res)}_{\alpha} V^{(D)}_{i, \alpha} \nonumber 
	+\sum_{\alpha} W^{\rm (self)}_{\alpha}\, \mathrm{MLP}\left( \{ V^{(D)}_{i, \alpha} \}; \, \bm\theta^{\rm (self)} \right) \\
	& & \qquad + \sum_{j \in \mathcal{N}(i)} \sum_\alpha {W}^{\rm (NN)}_{\alpha} \,\mathrm{MLP}\left( \{ V^{(D)}_{j, \alpha} \}; \, \bm\theta^{\rm (NN)} \right)
\end{eqnarray}
This construction maintains the same locality structure as the intermediate layers while allowing a flexible linear readout of both on-site and short-range correlated features to predict the scalar local energy.

\begin{figure}[t]
\centering
\includegraphics[width=0.99\columnwidth]{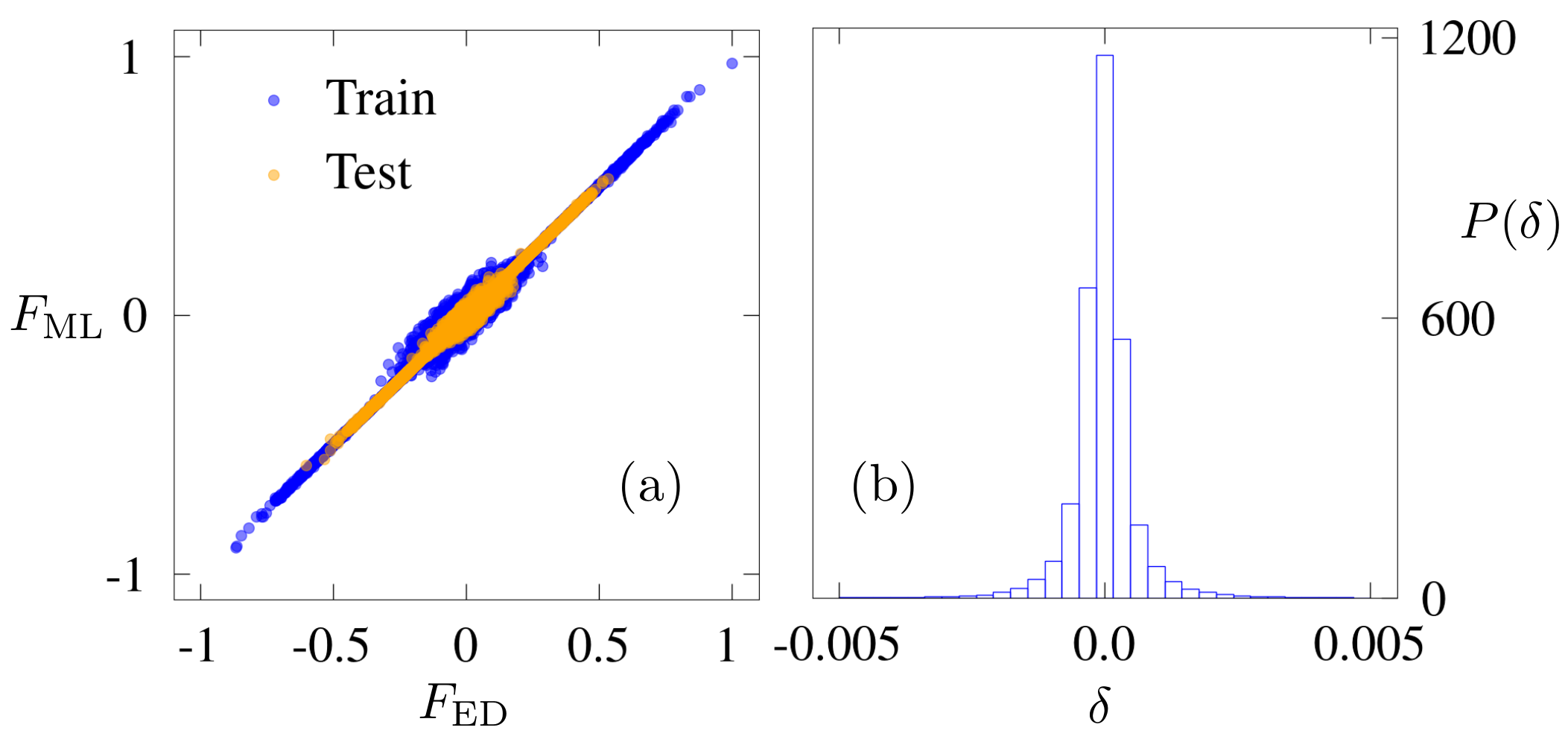}
\caption{Force prediction benchmark for the energy-based GNN model.
(a) Parity plot of machine-learning forces $F_{\mathrm{ML}}$---obtained from automatic differentiation of the predicted energy---versus exact-diagonalization values $F_{\mathrm{ED}}$ for training (blue) and test (orange) data. The close alignment along the diagonal indicates high accuracy and good generalization.
(b) Distribution of the force error $\delta = F_{\mathrm{ML}} - F_{\mathrm{ED}}$, sharply centered at zero, demonstrating unbiased and low-variance predictions.}
    \label{fig:MPNN-benchmark}
\end{figure}

We implement this GNN architecture in PyTorch~\cite{Paszke2019}; key hyperparameters are summarized in Table~\ref{tab:mpnn_parameters}. The network consists of multiple such message-passing blocks with progressively varying feature dimensions following a $1 \rightarrow 512 \rightarrow 1024 \rightarrow 512 \rightarrow 1$ channel structure. The final block maps the hidden representation back to a scalar local energy per node using the same message-passing and residual-update mechanism, but without an additional output nonlinearity. Overall, this design allows the model to encode local on-site effects and short-range correlations in a flexible yet physically controlled manner, while maintaining numerical stability and efficient training through residual connections and normalization.

Similar to the direct force-prediction scheme discussed in Sec.~\ref{sec:GNN}, successive message-passing layers progressively enlarge the effective receptive field of each node representation. After $\ell$ layers, the feature at site $i$ depends on lattice degrees of freedom within an $\ell$-step neighborhood, so that the network effectively constructs a nonlinear mapping from the local configuration $\mathcal{C}_i$ defined in Eq.~(\ref{eq:C_i}) to the local energy,
\begin{eqnarray}
	\epsilon_i = \varepsilon_{\rm GNN}\left(\mathcal{C}_i; \, \bm\theta \right).
\end{eqnarray}
The model thus provides a flexible yet explicitly local parameterization of the energy functional. Its validity relies on the locality principle, namely that the local energy density can be accurately approximated using information contained within a finite spatial neighborhood. The total energy is obtained through a simple extensive readout, $E = \sum_i \epsilon_i$, so that the GNN may equivalently be viewed as a mapping from the full configuration ${Q_i}$ to the scalar energy $E$. The forces are then computed by automatic differentiation, $F_i = -{\partial E}/{\partial Q_i}$, which guarantees consistency between energy and force predictions by construction. For training, we adopt the same force MSE loss defined in Eq.~(\ref{eq:loss-func}), allowing a direct comparison with the direct-force model.

Fig.~\ref{fig:MPNN-benchmark}(a) shows the force-prediction benchmark for the energy-based GNN. The predicted forces $F_{\rm ML}$ display an excellent linear correlation with the reference forces $F_{\rm ED}$ for both training (blue) and test (orange) sets, with no discernible systematic bias. The error distribution $P(\delta)$ in Fig.~\ref{fig:MPNN-benchmark}(b)  is sharply centered at zero and confined to a narrow window of order $10^{-3}$–$10^{-2}$, demonstrating that the model achieves high force accuracy while preserving strict energy–force consistency.
In comparison, the direct-force benchmark in Fig.~\ref{fig:GNN-benchmark} exhibits an even tighter collapse of $F_{\rm ML}$ onto $F_{\rm ED}$ and a visibly narrower error histogram, with fluctuations largely restricted to the $10^{-3}$ scale. This systematic improvement is expected, since the direct-force model optimizes the force field explicitly rather than through an intermediate energy functional.

\section{Large-scale GNN-based Langevin dynamics simulations}

\label{sec:coarsening}

\begin{figure*}[t]
\centering
\includegraphics[width=1.95\columnwidth]{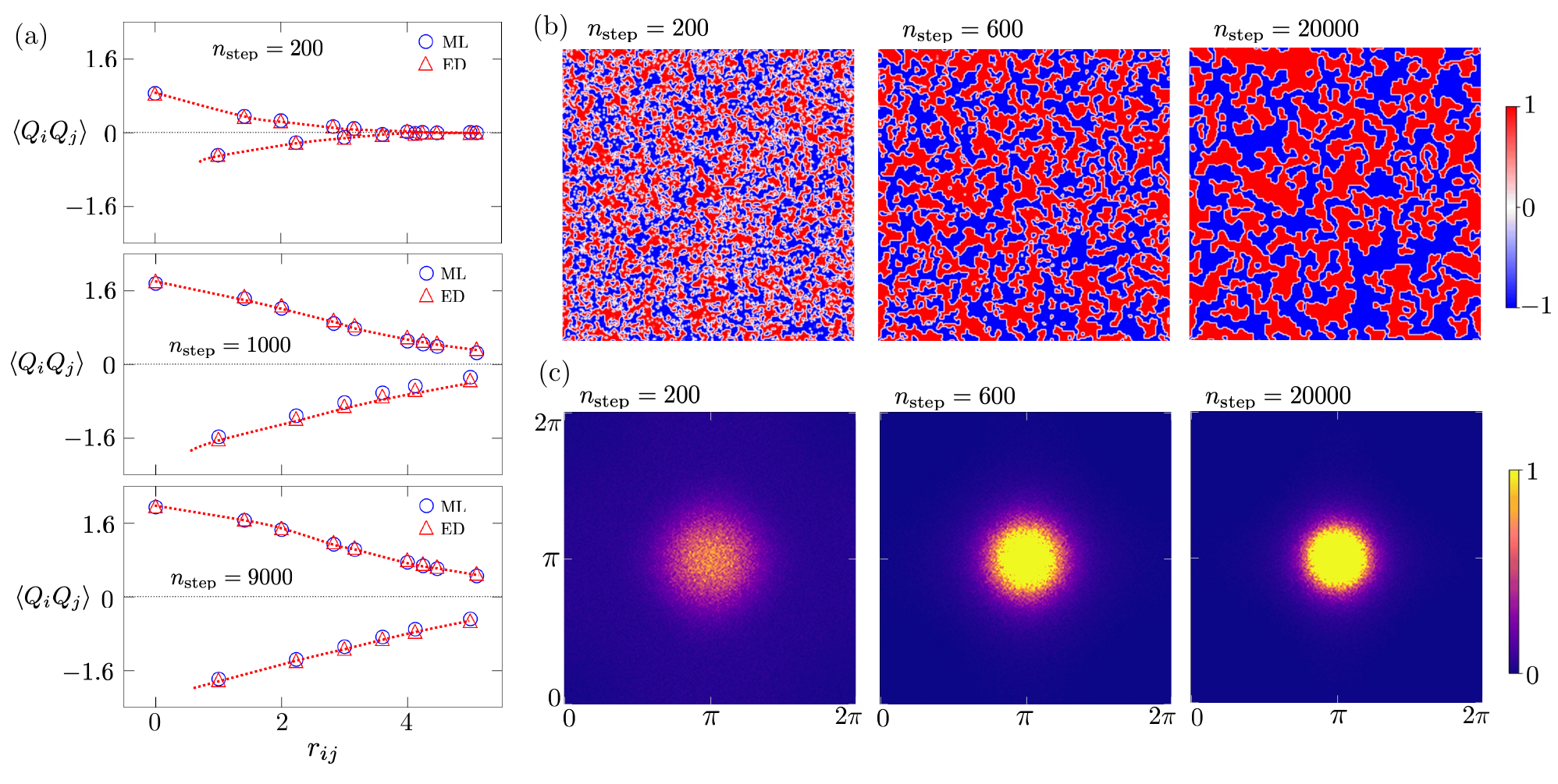}
\caption{Dynamical benchmark and large-scale quench simulations using the direct-force GNN model. (a) Comparison of the ensemble-averaged equal-time correlation function $\langle Q_i Q_j \rangle$ at three representative times after a thermal quench, obtained from exact diagonalization (ED)–based Langevin dynamics and machine-learning (ML)–based Langevin dynamics on a $40\times40$ lattice.   (b) Large-scale thermal quench simulations on a $200\times200$ lattice using the trained GNN force field, showing the real-space evolution and coarsening of charge-density-wave (CDW) domains at increasing time steps. The domain size grows systematically with time, illustrating curvature-driven coarsening dynamics. (c) Corresponding ensemble-averaged structure factor $S(\mathbf{q},t)$ at the same times as in (b).  }
    \label{fig:dyn-simulation}
\end{figure*}

As emphasized in Sec.~\ref{sec:intro}, scalability is a fundamental requirement for any practical force-field model. Across a broad range of applications---from {\em ab initio} molecular dynamics in quantum chemistry to semiclassical or adiabatic dynamics of correlated lattice Hamiltonians in condensed matter---the dominant computational cost arises from repeated evaluations of electronic-structure–derived forces. Because these evaluations must be performed at every time step for all degrees of freedom, the overall complexity must scale at most linearly with system size in order to reach physically relevant length and time scales. Without such favorable scaling, simulations of large molecules, extended materials, or mesoscale domain evolution quickly become intractable.

GNNs are naturally suited to this requirement. Their architecture is based on local message-passing operations, with learnable parameters shared across sites and edges, and acting only within finite neighborhoods. Consequently, the number of trainable parameters is independent of system size and—provided interactions are short-ranged—the computational cost per time step scales linearly with the number of degrees of freedom. This locality-preserving, size-extensive structure closely mirrors that of many physical Hamiltonians, making GNN-based force models well adapted for large-scale simulations.

In this section, we exploit this linear-scaling property to perform large-scale Langevin dynamics simulations of the Holstein model using the trained GNN force field. By replacing the costly ED-based force evaluation with on-the-fly GNN predictions, we access lattice sizes and time windows far beyond those attainable with conventional approaches. As a concrete example, we examine the nonequilibrium coarsening dynamics following a thermal quench and analyze the resulting dynamical scaling behavior of CDW domains.

We begin by benchmarking the GNN-driven dynamics against ED-based simulations. The trained GNN is integrated directly into the Langevin dynamics, with ED force evaluations replaced at each time step. We consider a thermal quench protocol in which an initially random configuration is coupled at $t=0$ to a heat bath at low temperatures. To present the results in a physically transparent form, we introduce natural dimensionless units. The fundamental time scale is set by the phonon oscillation period, $\tau_0 = 2\pi/\omega = 2\pi\sqrt{m/k}$. A characteristic lattice-distortion scale follows from balancing elastic and electron–phonon energies, $k Q_0^2 \sim g n Q_0$ with $n\sim1$, giving $Q_0 = g/k$. The corresponding momentum scale is $P_0 = m\omega Q_0$, and energies are measured in units of the nearest-neighbor hopping $t_{\rm nn}$.

The model is then characterized by a small set of dimensionless parameters. The electron–phonon coupling strength is quantified by $\lambda = g^2/k W$, where $W = 8 t_{\rm nn}$ is the electron bandwidth, which compares the typical electron–phonon energy to the electronic bandwidth. The ratio $\kappa/k$ controls the relative strength of the intersite distortive coupling, while the Langevin damping is parameterized by $\omega \tau_d$ with $\tau_d=\gamma/k$. Throughout this section we use $\lambda=1.5$, $\kappa/k=0.18$, weak damping $\omega\tau_d=0.09$, and a time step $\Delta t = 0.008\tau_0$.

Fig.~\ref{fig:dyn-simulation}(a) compares the GNN-driven Langevin dynamics with the reference ED-based results following a thermal quench. An initially random configuration is abruptly coupled at $t=0$ to a thermal bath at $T=0.1$. The figure displays the equal-time correlation function $C_{ij}=\langle Q_i Q_j\rangle$ at representative times ($n_{\rm step}=200$, 1000, and 9000) on a $40\times40$ lattice. The agreement between the GNN predictions (blue circles) and the ED results (red triangles) is essentially indistinguishable over all distances and time slices shown. This close correspondence indicates that the surrogate model does more than reproduce instantaneous forces at the pointwise level. It accurately captures the emergent spatiotemporal correlations that develop during the coupled electron–lattice relaxation process. In other words, the learned local force law remains dynamically consistent with the underlying many-body evolution over extended time scales.

Having established this dynamical benchmark on moderate system sizes, we next exploit the efficiency of the GNN to perform large-scale thermal-quench simulations on a $200\times200$ lattice—well beyond the practical limit of ED-based Langevin dynamics. Figs.~\ref{fig:dyn-simulation}(b) and (c) summarize the resulting coarsening behavior of the CDW order. Panel (b) presents real-space snapshots of the local CDW order parameter at $n_{\rm step}=200$, 600, and 20000. The local order parameter is defined as
\begin{align}
	\phi_i = \Bigl(n_i - \frac{1}{4}\sum_j\phantom{}^{'} n_j \Bigr) \exp\left({i \mathbf Q \cdot \mathbf r_i}\right), 
\end{align}
where the primed sum runs over nearest neighbors of site-$i$. This quantity measures the density contrast between a site and its immediate surroundings, modulated by the ordering wavevector $\mathbf Q = (\pi, \pi)$. A finite $\phi_i$ therefore signals local checkerboard charge modulation.

At early times, the system exhibits numerous small and irregular CDW domains nucleated from the random initial state. As the dynamics proceeds, these domains merge and coarsen; domain walls become smoother and the typical domain size increases. At long times, the lattice is partitioned into large regions of well-developed CDW order separated by relatively sharp interfaces. The evolution clearly reflects curvature-driven coarsening: smaller domains shrink and vanish, while larger ones expand, leading to a monotonic growth of the characteristic length scale. The ability to resolve such large-scale morphologies highlights the practical impact of the GNN-based approach, enabling simulations on lattices five times larger in linear dimension than the benchmark system.

The corresponding momentum-space evolution is shown in Fig.~\ref{fig:dyn-simulation}(c), where the structure factor
\begin{eqnarray}
	S(\mathbf k, t) = \left| \tilde{n}(\mathbf k, t) \right|^2,
\end{eqnarray}
is computed from the Fourier transform of the time-dependent density $n(\mathbf r_i,t)=n_i(t)$. As the system relaxes, spectral weight progressively concentrates at the ordering vector $\mathbf Q=(\pi,\pi)$. The initially broad and diffuse peak---reflecting short correlation lengths---gradually sharpens and increases in intensity, signaling the growth of long-range coherence and expanding CDW domains. The reciprocal-space signatures are fully consistent with the real-space coarsening patterns in panel (b), providing a complementary and quantitative characterization of the domain-growth dynamics enabled by the GNN-driven large-scale simulations.

\begin{figure}[t]
\centering
\includegraphics[width=0.9\columnwidth]{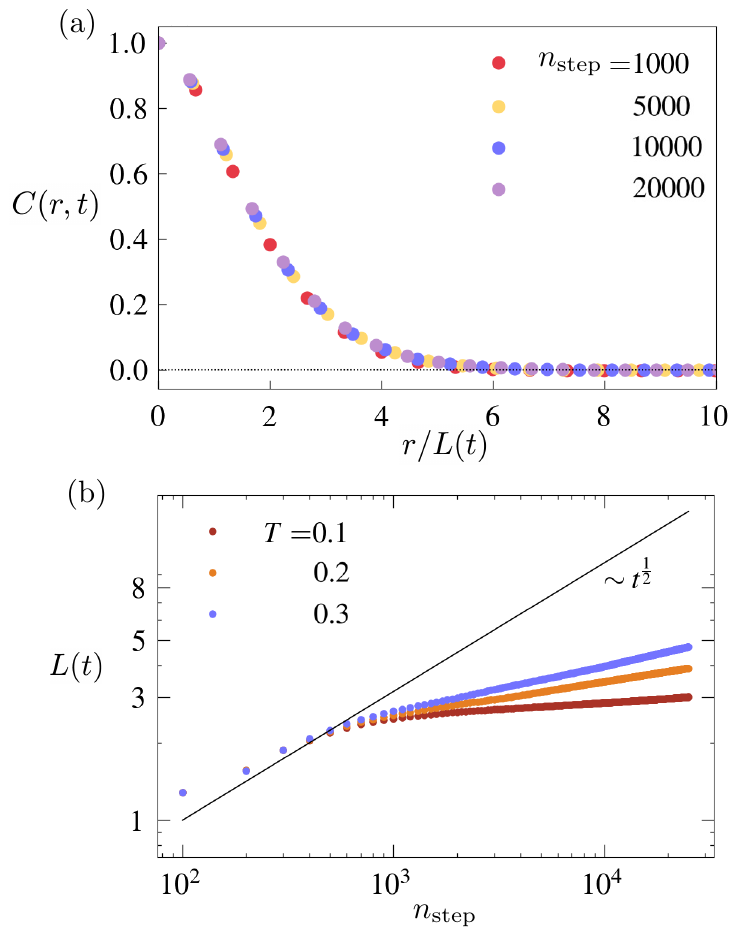}
\caption{(a) Ensemble-averaged equal-time correlation function $C(r,t)=\langle \phi_i \phi_j \rangle$ of the local CDW order parameter $\phi_i$, plotted versus the rescaled distance $r/L(t)$ at several times after the quench. The data collapse onto a single master curve demonstrates dynamical scaling governed by a single characteristic length scale $L(t)$. (b) Time evolution of the domain size (correlation length) $L(t)$ for three temperatures. The growth follows an approximate power law, $L(t)\sim t^{\alpha}$, with a temperature-dependent exponent $\alpha$ consistently smaller than the Allen-Cahn value $1/2$.}
    \label{fig:coarsening-Lt}
\end{figure}

To quantitatively characterize the coarsening dynamics, we evaluate the normalized equal-time correlation function of the local CDW order parameter,\begin{eqnarray}
	C(\mathbf r, t) = \frac{ \langle \phi(\mathbf r_0, t) \phi(\mathbf r_0 + \mathbf r, t) \rangle - \langle \phi(\mathbf r_0, t) \rangle^2 }
	{\langle \phi^2(\mathbf r_0, t) \rangle - \langle \phi(\mathbf r_0, t) \rangle^2}
\end{eqnarray}
Here $\langle \cdots \rangle$ denotes averaging both over the reference position $\mathbf r_0$ (spatial averaging within a single configuration) and over an ensemble of independent Langevin trajectories. The normalization ensures $C(\mathbf 0,t)=1$ and removes trivial shifts due to a nonzero spatial average of $\phi$, thereby isolating the growth of spatial correlations associated with domain formation. A time-dependent correlation length $L(t)$ is extracted from the decay of $C(\mathbf r,t)$ by defining it operationally through the half-maximum condition, $C(L(t), t) = 1/2$. This definition provides a robust measure of the typical CDW domain size and is insensitive to short-distance lattice-scale fluctuations.

Fig.~\ref{fig:coarsening-Lt}(a) presents the correlation functions at different times, plotted as a function of the rescaled distance $r/L(t)$. The excellent collapse of the data onto a single master curve demonstrates the emergence of dynamical scaling: at late times, the system becomes statistically self-similar, and the evolving domain morphology is governed by a single growing length scale $L(t)$. In other words, the correlation function adopts the scaling form
\begin{eqnarray}
	C(\mathbf r, t) = f\!\left( \frac{r}{L(t)} \right),
\end{eqnarray}
consistent with the standard scaling hypothesis for phase-ordering kinetics.

The extracted correlation length $L(t)$ is shown in Fig.~\ref{fig:coarsening-Lt}(b) as a function of time on a log–log scale for three different quench temperatures. The nearly linear behavior indicates approximate power-law growth,
\begin{eqnarray}
	L(t) \sim t^\alpha,
\end{eqnarray}
A fit to the late-time regime yields growth exponents $\alpha = 0.059$, $0.115$, and $0.155$ for increasing temperatures. Although $\alpha$ exhibits a clear temperature dependence, all extracted values are substantially smaller than the Allen–Cahn prediction $\alpha = 1/2$ expected for curvature-driven coarsening of a nonconserved scalar order parameter.

This sub–Allen–Cahn growth is consistent with previous descriptor-based ML studies of adiabatic Holstein dynamics~\cite{cheng23a}, indicating that it reflects intrinsic physics of the coupled electron–lattice system rather than a modeling artifact. In the standard Allen–Cahn picture~\cite{Allen1972}, domain-wall velocity is proportional to local curvature, leading to universal $t^{1/2}$ growth driven purely by interfacial tension. The marked suppression and temperature dependence of $\alpha$ observed here therefore point to additional kinetic constraints beyond simple curvature relaxation, likely involving thermally activated processes~\cite{Shore92,Corberi15}.

Such constraints become transparent in the strong-coupling limit. The local lattice displacement $Q_i$ develops two well-separated minima associated with $\langle \hat n_i\rangle=0$ and $1$, separated by an energy barrier $\Delta E\sim g^2/k$. Domain growth then requires activated transitions between these minima, increasingly suppressed at low temperatures. Furthermore, each local switch changes the electronic occupation by one; at half-filling, global electron-number conservation enforces correlated rearrangements rather than independent interface motion. Domain walls therefore advance through coordinated, energetically costly electron–lattice reconfigurations, naturally leading to slower-than-curvature-driven growth.

Resolving this unconventional scaling regime demands large system sizes and long time windows to suppress finite-size and transient effects. Such access is enabled here by the GNN-based force-field framework, which extends simulations far beyond the reach of direct ED approaches. The results thus highlight not only the fidelity of the learned force model, but also its utility in uncovering emergent mesoscale dynamics.

\section{Summary and Outlook}

\label{sec:summary}

In this work, we have developed a scalable and symmetry-consistent graph neural network (GNN) framework for force-field modeling of lattice Hamiltonians and demonstrated its effectiveness for the adiabatic dynamics of the semiclassical Holstein model. By treating the lattice as a highly symmetric graph, the GNN incorporates discrete translation and point-group symmetries directly through local message passing and parameter sharing, eliminating the need for explicitly constructed symmetry-adapted descriptors. We showed that both the direct-force and energy-based GNN formulations achieve high force accuracy when trained on exact-diagonalization data, while maintaining strict linear scaling with system size. Crucially, the learned local force functional generalizes seamlessly from small training lattices to substantially larger systems without retraining, demonstrating true size transferability.

Exploiting this scalability, we performed large-scale Langevin simulations of charge-density-wave (CDW) ordering following thermal quenches on lattices far beyond the reach of ED-based dynamics. The GNN-driven dynamics reproduce ED benchmarks at moderate system sizes and enable $200 \times 200$ simulations that reveal clear dynamical scaling behavior and sub-Allen--Cahn domain-growth exponents. These results establish that the GNN surrogate not only reproduces instantaneous forces with high fidelity, but also preserves the emergent spatiotemporal correlations governing nonequilibrium many-body evolution. More broadly, our findings demonstrate that message-passing architectures provide a conceptually transparent and computationally efficient route to symmetry-aware large-scale dynamical simulations of correlated lattice systems.

Looking forward, the GNN framework developed here admits a natural and systematic extension through integration with equivariant neural-network architectures, thereby enabling the treatment of lattice Hamiltonians with more intricate dynamical degrees of freedom. For systems involving multi-component lattice distortions—such as Jahn–Teller modes, vector phonons, or orbital pseudospins—equivariant layers can rigorously enforce the appropriate transformation laws under point-group operations and internal symmetry rotations, while the underlying message-passing structure preserves lattice locality, translation symmetry, and linear scalability. In this modular construction, the GNN encodes the spatial structure of the lattice, whereas equivariant components ensure the correct tensorial behavior of the dynamical variables and derived forces.

A similar strategy applies to itinerant magnetic systems with classical or semiclassical spin dynamics. By coupling lattice-based message passing with SO(3)-equivariant modules, one can guarantee rotational covariance of spin torques derived from a learned energy functional, thereby maintaining both symmetry consistency and energy conservation. More broadly, this unified GNN–equivariant framework is well suited for spin–fermion models, Kondo-lattice systems, multiferroics with coupled spin–lattice dynamics, and other correlated lattice Hamiltonians featuring intertwined internal and spatial symmetries. In this perspective, graph-based force-field models provide a general, extensible, and physically transparent platform for systematically bridging microscopic quantum Hamiltonians and large-scale nonequilibrium dynamics across a wide range of condensed-matter systems.

\begin{acknowledgments}
This work was supported by the US Department of Energy Basic Energy Sciences under Contract No. DE-SC0020330. The authors acknowledge Research Computing at The University of Virginia for providing computational resources and technical support that have contributed to the results reported within this publication.

\end{acknowledgments}
\bibliography{ref.bib}

\end{document}